\documentclass[12pt,a4paper]{article}

\usepackage{amssymb}
\usepackage{amsbsy}
\usepackage{amstext}
\usepackage{epsf}
\usepackage{rotating}
\usepackage[small]{caption}

\newcommand{\real}{I\kern-1.5mm{R}}       

\begin{document}
\thispagestyle{empty}
\unitlength 1mm 
\begin{picture}(0,0)(0,0)
\put(40,-52){\epsfxsize=5cm \epsfysize=5cm \epsffile{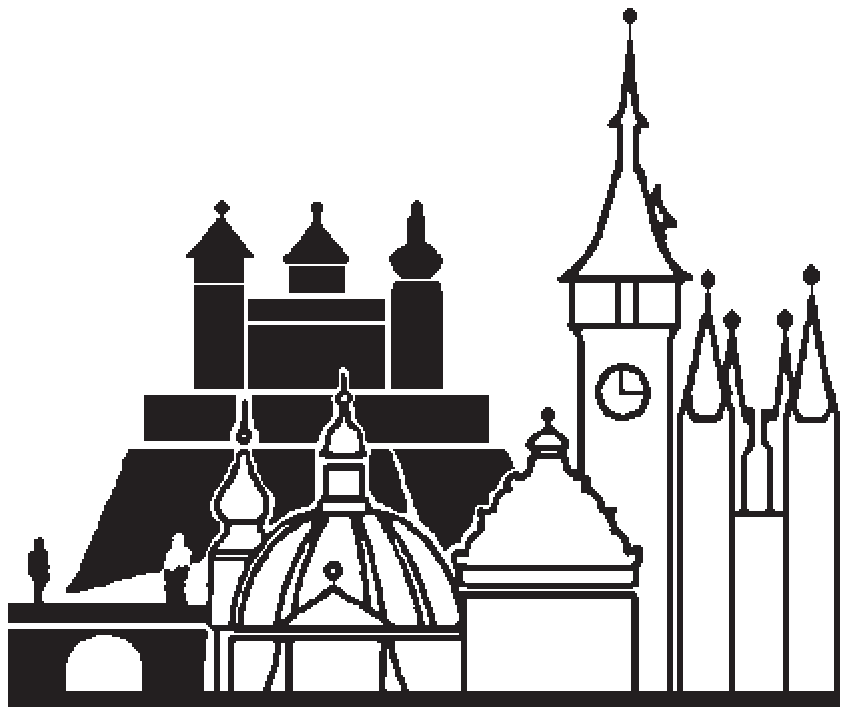}}
\end{picture}
\mbox{}
\vspace{-2cm}
\begin{center}
\large \sc
Universit\"at W\"urzburg\\[5cm]
Institut f\"ur Theoretische Physik\\[0.5cm]
\normalsize
Am Hubland, D--97074 W\"urzburg, Germany\\[2cm]
\begin{picture}(0,0)(0,0)
\put(0,0){\makebox(0,0)[cc]{$\diamondsuit$}}
\put(-28,0){\line(1,0){27}}
\put(1,0){\line(1,0){27}}
\end{picture}
\mbox{}\\[2cm]
\Large \sl
Phase Transitions of Neural Networks
\\[1cm]
\large
Wolfgang Kinzel \\[1cm]
Plenary talk for the MINERVA workshop on mesoscopics, fractals and
neural networks, Eilat, March 1997
\end{center}

\begin{picture}(0,0)(0,0)
\normalsize
\put(0,-50){\makebox[14cm]
{Ref.: WUE-ITP-97-005  \hfill
e-mail~: kinzel@physik.uni-wuerzburg.de}}
\put(0,-55){\makebox[14cm]
{ftp\hspace{1.1ex}: ftp.physik.uni-wuerzburg.de \hfill~}}
 \end{picture} 



\clearpage       
\newcommand     {\wvec}          {\mbox{\boldmath $w$\unboldmath}}
\newcommand     {\Svec}         {\mbox{\boldmath $S$\unboldmath}}
\newcommand     {\studout}       {\sigma}
\newcommand     {\p}[1]         {{ \left( {#1} \right) }}
       
\begin{center}        
{\Large\bf Phase transitions of neural networks}        
\end{center}        
\vspace{0,5cm}        
\begin{center}        
{\large Wolfgang Kinzel\\        
Institut f\"ur Theoretische Physik\\        
Universit\"at W\"urzburg, Am Hubland\\        
D-97074 W\"urzburg, Germany}        
\end{center}        
\vspace{1cm}        
  
\begin{abstract}  
The cooperative behaviour of interacting neurons and synapses is studied using models and   
methods from statistical physics. The competition between training error and entropy may lead   
to discontinuous properties of the neural network. This is demonstrated for a few examples:   
Perceptron, associative memory, learning from examples, generalization, multilayer networks,   
structure recognition, Bayesian estimate, on--line training, noise estimation and time series   
generation.  
\end{abstract}

\section{Introduction}        
        
Since about 15 years there exists a wave of interdisciplinary research activities under the topic        
''neural networks''. Neurobiologists, computer scientists, mathematicians, physicists,        
psychologists, and linguists are making a more or less common effort to understand the        
cooperative properties of a system of interacting neurons [Hertz et al 1991]. Meanwhile, the       
initial excitement        
and exaggerated promises have been replaced by practical research programs, but much has        
been achieved and many interesting and unexpected results have been obtained.        
        
The research on neural networks may be classified into three objectives:        
\begin{enumerate}        
\item Neurobiology: The material basis of our brain are about $10^{11}$ neurons, each of        
which is directly connected to about $10^3$ other ones via synapses. We know a lot about        
these single units and their anatomical and functional organization. However, we are still far        
away from understanding learning, association, memory, recognition and generalization on the        
basis of interacting neurons and their synaptic plasticity. It may be a philosophical problem        
whether mind, soul, creativity and consciousness can be understood by collective        
properties of a system of nerve cells. But there is a good chance to elucidate the basic        
properties of a real neural network by simple models.        
        
\item Computer science: There exists a variety of algorithms which use concepts from real        
neural networks. Simple units represent information and interact by synaptic weights. Such        
systems are trained by a set of examples. After the training phase, in which  the synaptic        
weights are adapted to the presented examples, the network is able to achieve a knowledge        
about the rule which produced the examples; it can generalize. These algorithms are called        
neural networks or neurocomputer; they are presently applied to a large variety of problems in        
engineering, science and economy. They have several advantages compared to standard        
approaches, and there is  hope to solve problems by neural networks which are too hard for        
methods of rule and data based algorithms of artificial intelligence.        
        
\item Physics: Neural networks definitely belong to the class of complex systems, which are        
characterized  by nonlinear dynamics,  feedback and  macroscopic properties emerging        
from a huge number of interacting units. In general, physics is interested in understanding such        
systems in terms of mathematical relations, scaling laws, phase transitions etc.        
\end{enumerate}          
      
In physics mathematical modelling of nature has been very successful. However, it is not clear        
at all whether such a complex system like the brain can be described by a mathematical        
language, by simple relations between macroscopic functions and microscopic mechanisms.       
        
On the other side, the full quantum mechanical description of an iron solid is not possible,        
either. Nevertheless one gains a lot of insight into the spontaneous magnetic ordering below a        
critical temperature if one studies the Ising model, which replaces the rather complex iron atom        
by a simple binary unit interacting with its neighbors. With this analogy it is definitely useful to        
investigate simple units, which model a few essential mechanisms of neurons and synapses, and       
to  study the cooperative behaviour of such interacting units. It is not obvious at all, whether       
such a system can store an infinite number of patterns with one set of synapses, learn from        
examples and generalize. Many questions can only be answered from a mathematical      
calculation.        
        
In this talk I want to emphasize the contribution of statistical
physics to the theory of neural computation. Using models and methods
from the physics of condensed matter one has been able to calculate
the properties of neural networks. This research program uses methods
developed already at the beginning of this century by L.~Boltzmann and
J.~W.~Gibbs. In 1975 S.~F.~Edwards and P.~W.~Anderson, S.~Kirkpatrick
and D.~Sherrington developed a theory of spin glasses [Fischer and
Hertz 1991] which was extended to neural networks by J.~J.~Hopfield
[1982]. The first analytic solution of attractor networks succeeded in
1985 [Amit et al 1987]. The statistical mechanics of learning from
examples was pioneered by the late E.~Gardner [1988]. These approaches
opened a new field of research, which produced a lot of interesting
results [for reviews see Watkin et al 1993, Opper and Kinzel 1996].
In view of the big challenge to understand the brain, the statistical
physics of neural networks will definitely survive over the next
century.
       
Long before the approach of statistical mechanics, mathematical models of neural networks        
have been investigated in detail with great success [Hertz et al 1991]. But in my talk I want to        
demonstrate        
on a few examples, to what extend the physics approach is able to ask questions and obtain        
results which are different from the approach of mathematics and computer science. In        
particular I think that only the methods of physics can calculate discontinuous and singular        
properties of infinitely large networks. Hence, in this paper I discuss examples from attractor       
networks, generalization, structure recognition,       
Bayesian estimate, on-line training, noise estimation and time series generation, which show        
discontinuous behavior as a function of model parameters or the size of the training set.       

This talk is not supposed to be a review. I apologize to all of my colleagues whose important 
contributions to the theory of neural networks are not mentioned.
       
\section{Perceptron}       
       
The simplest model of a neural network has already been introduced by Rosenblatt in 1960. It       
consists of an input layer of ''neurons" $S_i\; , i = 1, ..., N$, which take only binary        
values $S_i \in \{-1, +1\}$. The activity $S_i$ of each neuron is transmitted by ''synapses"        
$W_i \in \real $ to an output neuron $\sigma$ as shown in Fig.~1. The output reacts to the 
sign of the ''postsynaptic potential",             
\begin{equation}\label{eins}       
\sigma = \mbox{sign} \sum\limits^{N}_{i=1} W_i S_i =  \mbox{sign}\; \boldsymbol{W}       
\cdot       
\boldsymbol{S}       
\end{equation}

\begin{figure}
\begin{center}
\setlength{\unitlength}{1pt}
\begin{picture}(170,130)(0,0)
\put(0,0){\makebox(0,0)
{\includegraphics{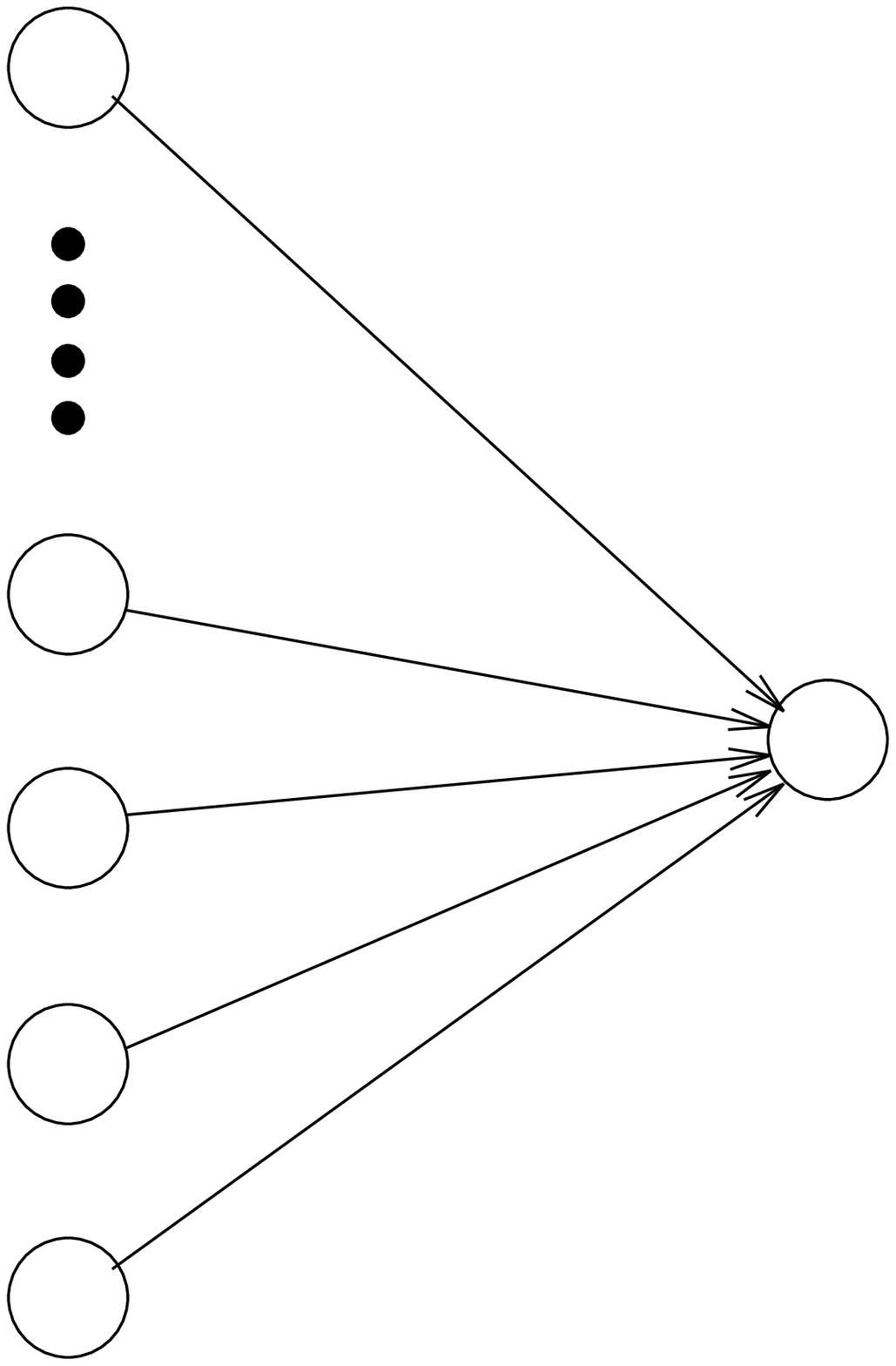}}}
\put(0,0){\makebox(0,0)
{\includegraphics{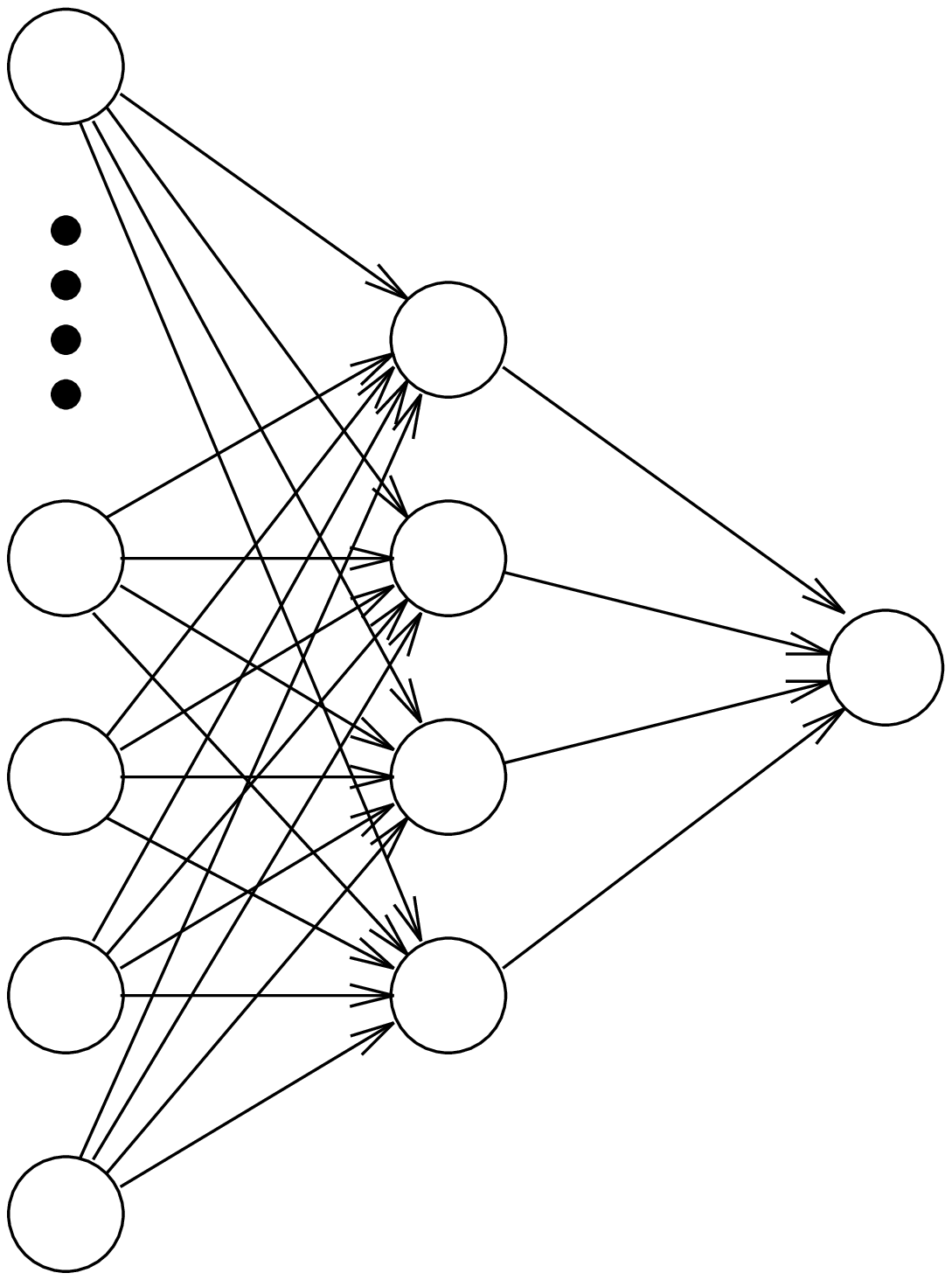}}}
\put(-115.00,130.00){\makebox(0,0)[lc]{$S_1$}}
\put(-85.00,130.00){\makebox(0,0)[lc]{$S_2$}}
\put(45.00,130.00){\makebox(0,0)[lc]{$S_N$}}
\put(102.00,130.00){\makebox(0,0)[lc]{$S_1$}}
\put(252.00,130.00){\makebox(0,0)[lc]{$S_N$}}

\put(-100.00,75.00){\makebox(0,0)[lc]{$w_1$}}
\put(-5.00,75.00){\makebox(0,0)[lc]{$w_N$}}

\put(105.00,85.00){\makebox(0,0)[lc]{$\wvec_1$}}
\put(248.00,85.00){\makebox(0,0)[lc]{$\wvec_K$}}
\put(132.00,40.00){\makebox(0,0)[lc]{$1$}}
\put(212.00,40.00){\makebox(0,0)[lc]{$1$}}

\put(-10.00,15.00){\makebox(0,0)[lc]{$\studout=g\p{\wvec\cdot\Svec}$}}
\put(191.00,5.00){\makebox(0,0)[lc]{$\studout=f\p{\sum_{i=1}^K g\p{\wvec_i\cdot\Svec}}$}}
\end{picture}
\end{center}
\caption{Architecture of the perceptron (left) and the committee machine(right). }
\end{figure}
       
In the training phase this network, which was called ''perceptron", receives a set of training        
examples $(\sigma^{\nu}, \boldsymbol{S}^{\nu}), \; \nu = 1, ... ,\alpha N$. It changes its       
weight        
$W_i$ such that a maximum number of examples is correctly mapped by Eq.~(\ref{eins}). A        
simple algorithm has been investigated by Rosenblatt (see Hertz et al 1991): It presents the       
examples in        
an arbitrary sequence. If an example is not correctly classified, that is if $\boldsymbol{W} (t) \cdot        
\boldsymbol{S}^{\nu} \sigma^{\nu} < 0$, then       
\begin{equation}\label{zwei}       
\boldsymbol{W}(t+1) = \boldsymbol{W}(t) + \frac{1}{N}\; \boldsymbol{S}^{\nu} \cdot        
\sigma^{\nu}       
\end{equation}       
There exists a convergence proof for this algorithm: If the examples can be mapped correctly        
by any perceptron, Eq.~(\ref{eins}) with weights $\boldsymbol{W}^*$, then the perceptron       
rule        
Eq.~(\ref{zwei}) finds a solution, i.~e.~the algorithm stops.       
       
The Rosenblatt training rule stems from neurobiology, as proposed by D.~Hebb in 1949:        
Each synapse reacts to the neuronal activities at its two ends. Here we need an additional        
influence of the postsynaptic potential.       
       
The perceptron implements a linear separable Boolean function, which
has an interesting geometrical interpretation: $\boldsymbol{W} \cdot
\boldsymbol{S} = 0$ defines a hyperplane in an $N$--dimensional space
of inputs $\boldsymbol{S}$, the weight vector $\boldsymbol{W}$ is
normal to this plane. On the side of the vector $\boldsymbol{W}$ the
perceptron classifies each input $\boldsymbol{S}$ to $\sigma = + 1$
(black, correct, ...), on the other side the label is $\sigma =
-1$ (white, wrong, ...) see Fig.~2.\\
\begin{figure}
\centerline{
\epsfysize=5cm
\epsffile{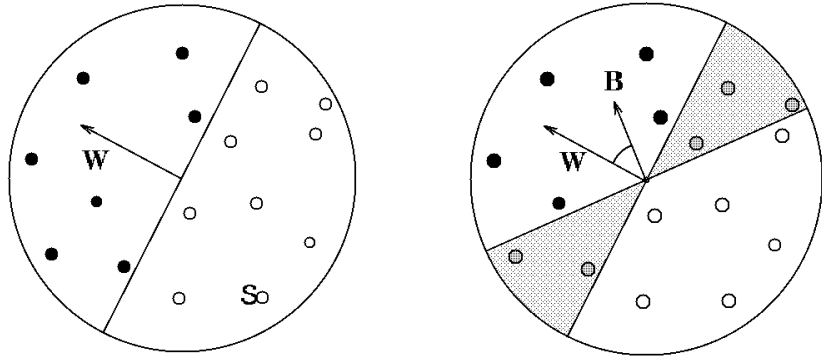}
}
\caption{  Space of input vectors $\boldsymbol{S}$. The weight vector  
$\boldsymbol{W}$ of the perceptron defines a hyperplane in the $N$--dimensional space,  
which separates the labels $\sigma$ of the input vectors $\boldsymbol{S}$. In the shaded  
region the labels of teacher $\boldsymbol{B}$ and student perceptron $\boldsymbol{W}$ are  
different. }
\end{figure}
       
\noindent Now we consider a set of  $\alpha N$ many points $\boldsymbol{S}^{\nu}$ in $N$        
dimensions. How many sets of labels $\{\sigma^{\nu}\}$ can be represented by any perceptron?        
Surprisingly this problem which is important for the theory of neural computation was already        
solved by the Swiss mathematician Ludwig Schl\"afli in the last century [Schl\"afli 1852]. If       
any subset of        
$N$ points is linearly independent, then the number $C$ of possible sets of labels        
$\{\sigma^{\nu}\}$ is given by       
\begin{equation}\label{drei}       
C = 2 \sum\limits^{N-1}_{i=0}  {\alpha N-1 \choose i} .       
\end{equation}       
For $\alpha <1$ all labels can be produced by a perceptron, i.~e.~
$C=2^{\alpha N}$. For $\alpha < 2$ there is a large fraction of labels
which can be separated by a hyperplane.  For $\alpha >2$ only a tiny
fraction of patterns can be stored, this fraction disappears for $N
\rightarrow \infty$.  This result has consequences for the associative
memory which will be discussed in the following section: In the limit
of $N \rightarrow \infty$ a network with $N$ neurons can store up to
$2N$ random patterns.
\begin{figure}
\centerline{
\epsfysize=5cm
\epsffile{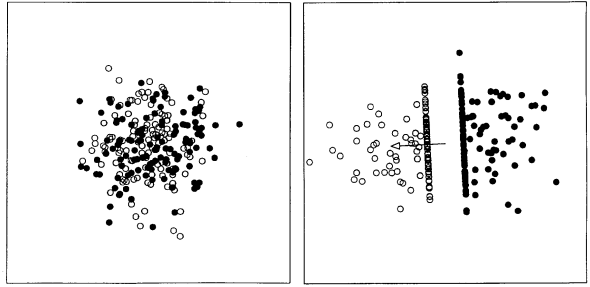}
}
\caption{ Two dimensional projections of 250 points in 200 dimensions. The points are labelled  
randomly. The perceptron algorithm finds a hyperplane which separates different labels. }
\end{figure}

The geometry of this result is shown in Fig.~3. 250 points are located in a 200 dimensional        
space and randomly colored black or white with equal probability. Now we are        
moving in the space of points and would like to find a view where black and white is clearly        
separated by a single gap. From Eq.~(\ref{drei}) we find with $N=200$ and $\alpha N=250:        
\hspace{0.5cm} C/2^{250} \simeq 1-4\cdot 10^{-23}$; that means for random labels it is        
almost sure that one can find such a view. In fact the Rosenblatt algorithm, Eq.~(\ref{zwei}),        
found the solution shown in Fig.~3.       
       
\section{Attractor networks}       
The perceptron  is the ''atom" of all neural networks. Many of such elementary units can be        
composed to a large and complex network. Here we consider an attractor network which        
consists of $N$ neurons $S_i$ as before. But now each element $S_i$ is connected to any other        
element $S_j$ by a coupling $W_{ij} \in \real$, as illustrated in Fig.~4.
\begin{figure}
\centerline{
\epsfysize=5cm
\epsffile{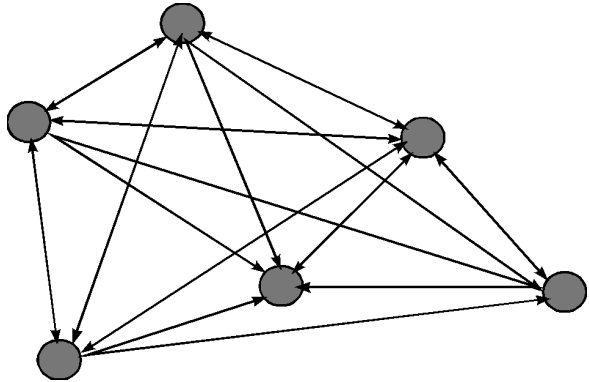}
}
\caption{ A set of six perceptrons is connected to form an attractor network. }
\end{figure}

We want to store $\alpha N$ many patterns $S^{\nu}_{i} \in \{-1, +1\}\;  ; i=1, .., N\; ; \nu =1,       
..., \alpha N$. If we use the Rosenblatt rule, Eq.~(\ref{zwei}), without the additional 
condition, we obtain the Hebbian couplings       
\begin{equation}\label{vier}       
W_{ij} = \frac{1}{N} \sum_{\nu} S^{\nu}_{i} S^{\nu}_{j} \,\,\,\,\,\,\,\, (i \neq j).       
\end{equation}       
Since each input $S_j$ is output of a perceptron with weights $W_{jk}$, we can define a        
dynamics of the configuration of neurons, $\boldsymbol{S}$. For instance, for each neuron        
$S_i$ we consider the local field       
\begin{equation}\label{fuenf}       
h_i = \sum_j W_{ij} \; S_j (t)       
\end{equation}       
where $t$ is a discrete time index. Now we define a stochastic dynamics by the probability $P$        
to find neuron $S_i$ in the state $S \in \{+1, -1\}$ in the next time step $t+1$:       
\begin{equation}\label{sechs}       
P[S_i (t+1) =S] = \frac{e^{\beta h_{i}S}}{2 \cosh (\beta h_{i}S)}       
\end{equation}       
$\beta$ is a parameter which measures the noise level of the dynamics. For $\beta \rightarrow        
\infty$ we obtain the noiseless deterministic equation       
\begin{equation}\label{sieben}       
S_i (t+1) = \mbox{sign} \sum_j W_{ij} S_j (t).       
\end{equation}       
This model was introduced by Hopfield [1982]. He noticed that the dynamics of the neurons is        
nothing else than the usual Monte Carlo procedure to obtain thermal equilibrium. Since the        
couplings are symmetric, $W_{ij}= W_{ji}$, the stationary state is given by a Boltzmann        
distribution        
\begin{equation}\label{acht}       
P(\boldsymbol{S}) = \exp (-\beta H(\boldsymbol{S})) / Z       
\end{equation}       
with a Hamiltonian       
\begin{equation}\label{neun}       
H(\boldsymbol{S}) = -\frac{1}{2} \sum_{i \neq j} W_{ij} S_{i}S_{j}       
\end{equation}       
This is the main advantage of equlibrium statistical mechanics: The dynamics        
$\boldsymbol{S} (t)$ is replaced by a summation over all possible states $\boldsymbol{S}$.        
Instead of solving a system of $N$ strongly coupled nonlinear equations, one has to        
calculate the partition sum       
\begin{equation}\label{zehn}       
Z = \sum_{\{\boldsymbol{S}\}} \exp \left[-\frac{\beta}{2} \sum_{i\neq j} W_{ij} S_i S_j \right].       
\end{equation}       
In the thermodynamic limit of infinitely many neurons, $N \rightarrow \infty$, and infinitely        
many patterns, $\alpha = $const., the partition sum $Z$ was solved exactly by Amit et al       
[1987] using        
the replica method. There are two main steps in the calculation:       
\begin{enumerate}       
\item The free energy $f= - \ln Z/ \beta$ is averaged over all possible sets of patterns        
$\{\boldsymbol{S}^{\nu}\}$. It can be shown that the average value gives the same results as        
the value $f$ for a single, randomly chosen set of patterns. Hence, this calculation yields results 
for a typical situation.       
\item The sum over $2^N$ states in $\ln Z$ is performed for fixed order parameters. The        
minimum of $f$ as a function of these quantities yields their physical values, which describe the        
stationary state. Hence, the complex system of interacting neurons is described exactly by a        
few order parameters.       
\end{enumerate}       
       
The first step is done using the replica method:       
\begin{equation}\label{elf}       
\langle \ln Z \rangle_{\{\boldsymbol{S}^{\nu}\}} = \lim_{n \rightarrow 0}        
\frac{\partial}{\partial n}  \langle Z^n \rangle_{\{\boldsymbol{S}^{\nu}\}}.       
\end{equation}       
In the second step one is interested in the overlap between the state
$\boldsymbol{S}$ and one of the patterns $\boldsymbol{S}^{\nu}$. Let
us assume that the first pattern $\boldsymbol{S}^1$ has the form of an
''{\bf A}" as shown in Fig.~4 with $N=400$. The other 31 patterns
consist of random bits. If the initial state $\boldsymbol{S}(0)$ has
an overlap to the first pattern, for instance if it is the noisy
''{\bf A}" of Fig.~5, then after a few steps given by
Eq.~(\ref{sieben}) the network relaxes to the stored information more
or less completely.
\begin{figure}
\centerline{
\epsfysize=4cm
\epsffile{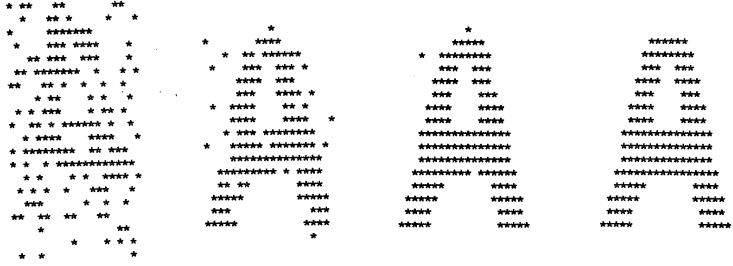}
}
\caption{ An initial state of an attractor network relaxes to one of the stored patterns. 
(From Kinzel 1985)}
\end{figure}
       
The statistical mechanics gives information about the possible final states of the dynamics.        
Here we are interested in the overlap after the relaxation:       
\begin{equation}\label{zwoelf}       
m_A = \frac{1}{N} \boldsymbol{S}\; \boldsymbol{S}^A.       
\end{equation}       
It turns out that one obtains this order parameter if one calculates the free energy. The        
overlaps       
\[       
m_{\nu} = \frac{1}{N} \; \boldsymbol{S} \cdot  \boldsymbol{S}^{\nu}       
\]       
to the other 31 patterns are of the order of $1/ \sqrt{N}$. However, their sum $r$ is an        
additional order parameter       
\[       
r = \frac{1}{\alpha} \sum\limits^{\alpha N}_{\nu =2} m_{\nu}^2       
\]       
$r$ measures the fluctuation of the final state to the rest of the patterns. Finally there is an        
order parameter $q$ which measures the complexity of the space of possible solutions        
$\boldsymbol{S}$. Like in the theory of spin glasses, it signals an additional order of the        
stationary states which has no simple interpretation [Fischer and Hertz 1991].       
       
The theory of attractor networks has close similarities to the theory of an Ising ferromagnet        
with infinite range interactions. In both cases energy and entropy can be expressed in terms of        
order parameters. For the ferromagnet one obtains an implicit equation for the spontaneous        
magnetic order $m$ [Yeomans 1992]       
\begin{equation}\label{dreizehn}       
m = \tanh \beta Jm \, \, .       
\end{equation}       
For the attractor network one finds       
\begin{equation}\label{vierzehn}       
m_A = \langle \tanh (\beta m_A + \beta \sqrt{\alpha r}z )\rangle_z      
\end{equation}       
where the average is performed over a Gaussian distributed quantity $z$. There are additional       
equations for $r$ and $q$. Hence, compared to the ferromagnet, the $\alpha N-1$ patterns add        
a noise term to the local fields.       
\begin{figure}
\centerline{
\epsfysize=6cm
\epsffile{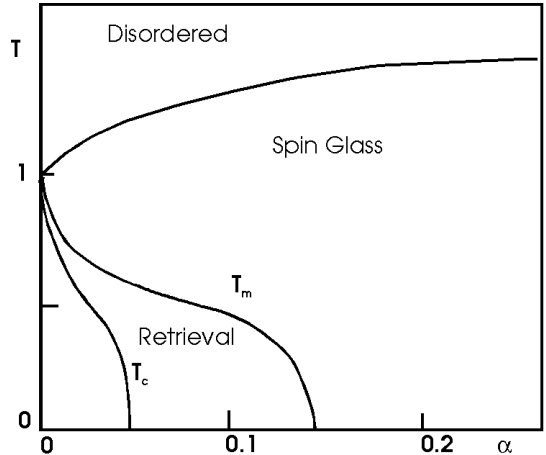}
}
\caption{ Schematic phase diagram of the Hopfield model. (From Amit et al, 1987) }
\end{figure}
       
Fig.~6 shows the result of the analytic calculation [Amit et al 1987]. In the
noise--load plane one obtains several phases, which are well separated
in the thermodynamic limit. For strong noise, $T = 1/\beta >1$, or for
high load, $\alpha > 0.14$, the network cannot recognize the stored
patterns at all. Nevertheless there is a spin glass order for low
noise $T < T_{g}$ with $q>0$. Only for $T <T_m (\alpha)$ the network
can relax to final states which have an overlap to one of the stored
patterns. This overlap jumps discontinuously to zero at $T_m$. For $T
<T_c (\alpha)$ this retrieval state has the lowest free energy,
i.~e.~it is thermodynamically stable. Note that also for the
deterministic dynamics $(T=0)$ there is a discontinuous drop to zero
retrieval at $\alpha = \alpha_c \simeq 0.14$. For $\alpha < \alpha_c$
the network restores stored information very well, for $\alpha >
\alpha_c$ the network relaxes to final states which have nothing in
common with the stored patterns. 

According to Schl\"afli there are couplings with a storage capacity
of $\alpha_c = 2$, but these interactions are not symmetric and one
cannot apply statistical mechanics to solve the corresponding
attractor network. A network with $\alpha_c = 1$ has been analysed by
Kanter and Sompolinsky [1987].
       
In summary, the attractor network functions as an associative memory. It is a distributed        
memory, since all patterns are stored in all couplings. It is content addressable, since a state        
with a partial information relaxes to the complete one. Even with a stochastic dynamics it        
performs well, if the noise level $T$ and storage capacity $\alpha$ are not too high. There is a        
sharp, discontinuous transition between good and zero performance. Processing of information emerges
as a cooperative effect from a large number of simple interacting units.      
       
\section{Generalization}       
We have seen how a network of mutually interacting perceptrons can work as an        
associative memory. But already the simple perceptron itself has interesting properties. It can        
learn from examples and recognize an unknown rule.       
       
Consider a perceptron with a weight vector $\boldsymbol{W}$, as in
Eq.~(\ref{eins}). We will call this perceptron the ''student". It
obtains a set of examples $(\sigma_{B}^{\nu}, \boldsymbol{S}^{\nu} ),
\nu = 1, ..., \alpha N$ from a ''teacher". In the simplest case the
teacher is another perceptron with a weight vector $\boldsymbol{B}$.
To what extent can the student gain information about the vector
$\boldsymbol{B}$ if the only available information is the set of
$\alpha N$ many examples? The patterns $\boldsymbol{S}^{\nu}$ are
selected randomly and $\sigma^{\nu}_{B}$ is the output of the teacher,
\begin{equation}\label{fuenfzehn}       
\sigma^{\nu}_{B} = \mbox{sign} \; \boldsymbol{B} \cdot \boldsymbol{S}^{\nu}       
\end{equation}       
As before we are interested in the limit $N \rightarrow \infty$ and $\alpha =$constant. \\       
We have to consider two processes:       
\begin{enumerate}       
\item The training phase:\\       
The student network is trained by use of the examples, it tries to decrease the training error       
\begin{equation}\label{sechzehn}       
\varepsilon_t (\boldsymbol{W}) = \sum\limits^{\alpha N}_{\nu =1} \;\; \Theta [-       
\sigma^{\nu}        
\cdot \sigma^{\nu}_{B} ]       
\end{equation}       
$\Theta(x)$ is the step function, it is zero if the student  reproduces the example        
$\boldsymbol{S}^{\nu}$ correctly.       
       
\item  The test phase:\\       
Now the student receives an input $\boldsymbol{S}$ which  has not been presented before. It gives       
the        
answer $\sigma = \mbox{sign}\;  \boldsymbol{W} \cdot \boldsymbol{S}$, which may be       
different from the        
answer by the teacher, $\sigma_B = \mbox{sign}\; \boldsymbol{B} \cdot \boldsymbol{S}$. The       
probability        
of disagreement or the generalization error is defined by an average over all possible input        
vectors $\boldsymbol{S}$:       
\begin{equation}\label{siebzehn}       
\varepsilon_g = \langle \Theta [- \sigma \; \sigma_B] \rangle_{\boldsymbol{S}}        
\end{equation}       
From Fig.~2 one can see that $\varepsilon_g$ is determined by the angle between the        
weight vectors of the student and the teacher perceptron       
\begin{equation}\label{achtzehn}       
\varepsilon_g = \frac{1}{\pi} \arccos \frac{\boldsymbol{B} \cdot        
\boldsymbol{W}}{|\boldsymbol{B}| \, |\boldsymbol{W}|}       
\end{equation}       
\end{enumerate}       
Training and generalization of the perceptron has been studied in
detail using methods of statistical mechanics [see e.g. Watkin et al
1993, Opper and Kinzel 1996] on a simple scenario where the weights
are restricted to binary values,
$W_i \in \{+1, -1\}$ and $B_i \in \{+1, -1\}$. The student perceptron is
trained by a stochastic algorithm, for instance by a Monte Carlo
procedure similar to Eq.~(\ref{sechs}). But now we have a stochastic
dynamics of the synaptic weights $W_i$ instead of the neurons $S_i$,
which leads to a thermal equilibrium given by
\begin{equation}\label{neunzehn}       
P(\boldsymbol{W}) = \exp [-\beta \; \varepsilon_t (\boldsymbol{W})]  /Z     
\end{equation}       
As before we do not have to solve the complex nonlinear dynamics of the weights        
$\boldsymbol{W}(t)$ but rather calculate the partition sum       
\begin{equation}\label{zwanzig}       
Z = \sum\limits_{\{\boldsymbol{W}\}} \;\; \exp [-\beta \; \varepsilon_t (\boldsymbol{W})]       
\end{equation}       
Again we have to perform two steps:       
\begin{enumerate}       
\item Average $\ln Z$ over all possible sets of examples $\{\boldsymbol{S}^{\nu}\}$.       
\item Evaluate the sum of $2^N$ states $\boldsymbol{W}$ by introducing order parameters.       
\end{enumerate}       
In the limit of large noise, $\beta \rightarrow 0$, the calculation turns out to be very easy        
[Seung et al 1992]:       
The only order parameter is the overlap $R$ between student and teacher,       
\begin{equation}\label{einundzwanzig}       
R = \frac{1}{N} \; \boldsymbol{B} \cdot \boldsymbol{W}       
\end{equation}       
The training error of Eq.~(\ref{zwanzig}) can be replaced by the generalization error       
\begin{equation}\label{zweiundzwanzig}       
\alpha \; \varepsilon_g = \frac{\alpha}{\pi} \arccos R       
\end{equation}       
and the entropy is the well known mixture entropy of binary variables       
\begin{equation}\label{dreiundzwanzig}       
S(R) = \frac{1}{2} [ (1+R) \; \ln(1+R) + (1-R)\; \ln(1-R)] + \ln2        
\end{equation}       
$R$ is determined by the minimum of the free energy       
\begin{equation}\label{dreiundzwanziga}       
f(R) = \alpha \varepsilon_g ( R ) - T S( R)       
\end{equation}       
Note that the product $\beta \alpha$ in the limit $\beta \rightarrow 0$ is the only parameter of        
the model, hence a large noise has to be compensated by a large number of examples.       
\begin{figure}
\centerline{
\epsfysize=5cm
\epsffile{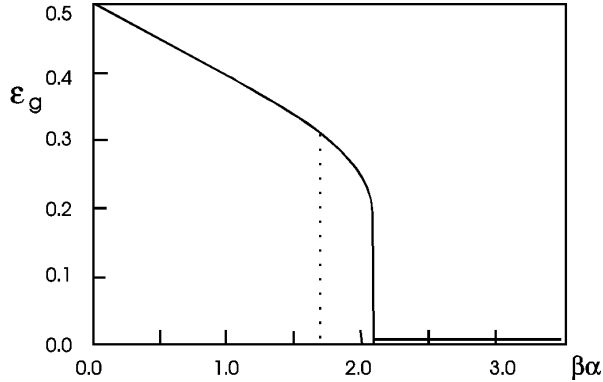}
}
\caption{ Generalization error $\varepsilon_g$ as a function of the size $\alpha$ of the  
training set (schematic). The binary perceptron is trained stochastically for large noise $(\beta  
\rightarrow 0)$. The dotted line describes the first order phase transition to perfect  
generalization in thermal equilibrium. The solid line extends to a metastable state. (From Seung  
et al 1992) }
\end{figure}
       
One minimum of $f$ is always $R=1$, i.~e.~, the student perfectly recognizes the teacher.        
However, for $\beta\alpha < 2.08$, the system has an additional minimum at $R<1$ which is        
the global one for $\beta\alpha < 1.69$. Fig.~7 shows the generalization error as a function        
of the fraction of learned patterns. There is a discontinuous transition from poor to perfect        
generalization, similar to a first order phase transition in solid state physics. Both of the        
transitions are characterized by metastable states and hysteresis loops. This process of        
sudden recognition appears even for a noisy training algorithm. A replica calculation shows        
that the transition qualitatively extends to zero noise $T=0$. [Seung et al, 1992]      
       
\section{Multilayer networks}       
We have already seen how an attractor network can be built from many perceptrons.
Another interesting system which can be constructed from many elementary units is a        
multilayer network, shown in Fig.~1. It consists of several layers of synaptic weights which        
map the information coded in the neurons from top to bottom. It is important that such        
networks can realize any function, if the number of hidden units (neurons in the interior layers)        
is large enough.       
       
The simplest multilayer network is a committee machine. It consists of $N$ input units, $K$        
hidden units, $K$ weight vectors $\boldsymbol{W}_i, i=1, ..., K$ and one output unit       
$\sigma$.        
The weights of the second layer have the value $+1$, that means, that the output bit $\sigma$        
is given by the majority of the $K$ perceptrons (= members of the committee) in the first layer,       
\begin{equation}\label{vierundzwanzig}       
\sigma = \mbox{sign} \left[ \sum\limits^K_{i=1} \;  \mbox{sign} \; \boldsymbol{W}_i \;       
\boldsymbol{S} \right]       
\end{equation}       
This network is trained from a set of examples $(\sigma^{\nu}_B, \boldsymbol{S}^{\nu}), \nu       
=        
1, ..., \alpha N$. Note that the opinion of the majority is trained, not the opinion of each        
member of the committee!       
       
Here we consider the case, where the student is a committee machine with $K=3$ members.        
The teacher is a simple perceptron with single weight vector $\boldsymbol{B}$. All the weights       
are        
assumed to be binary, $W_{ik}, B_i \, \in \, \{+1, -1\}$. To what extent can a complex        
network gain information about a simple rule from a set of examples?       
       
In analogy to the previous section we consider a stochastic training
algorithm. The training error $\varepsilon_t$ is the ''energy" of the
Gibbsian probability which describes the stationary state of the
stochastic algorithm. In the limit of high noise there are several
order parameters, which determine the energy, entropy and
generalization error. Firstly, there are the overlaps between the
members of the committee and the teacher,
\begin{equation}\label{fuenfundzwanzig}       
R_i = \frac{1}{N} \boldsymbol{W}_i \cdot \boldsymbol{B} \;\;\; (i = 1, 2,3)\;\;.       
\end{equation}       
Secondly, the weight vectors of the committee have a mutual overlap       
\begin{equation}\label{sechsundzwanzig}       
Q_{ij} = \frac{1}{N} \boldsymbol{W}_i \boldsymbol{W}_j       
\end{equation}       
Their physical values are determined by minimizing the corresponding free energy.\\
Fig.~8 shows the result of the analytic calculation [Schwarze et al
1992]. For a small size of the training set ($\beta\alpha$ small) the
members of the committee react symmetrically, $R_1 = R_2 = R_3 < 1$
and the generalization error decreases continuously with $\alpha$.
By increasing the size of the training set, suddenly one of the members recognizes
the teacher perfectly, $R_1 =1, R_2 =R_3 <1$, and the error jumps to a
low value. Further increase of  $\beta\alpha$ leads to another
discontinuous transition to perfect recognition of the majority. Since
the majority vote is already determined by two members, the highest
entropy is achieved for $R_1 = R_2 =1$ and $R_3 = 0$. 
\begin{figure}
\centerline{
\epsfysize=4cm
\epsffile{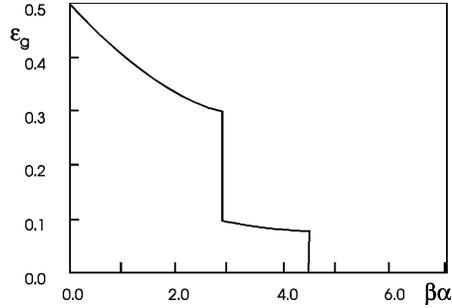}
}
\caption{ Generalization error $\varepsilon_g$ as a function of the size of the training set  
(schematic). A committee machine with binary weights and three hidden units is trained to a set  
of examples given by a binary perceptron. (From Schwarze et al 1992) }
\end{figure}

Here again the
competition between energy and entropy leads to an interesting
discontinuous behavior of the generalization ability. Such sharp
transitions, which occur for infinitely large networks, only $(N
\rightarrow \infty)$, are not obvious. One needs the tools of
statistical mechanics to find and describe them.
  
\section{Parity machine}  
  
Now we want to discuss another multilayer network, the parity machine with a tree   
architecture shown in Fig.~9. It consists of $N$ input and $K$ hidden units. The input units 
are grouped into   
$K$ parts with $N/K$ neurons each. Each part is input of a perceptron with weights   
$\boldsymbol{W}_i \; (i=1,...,K)$. The output of the whole network is given by the parity of   
the outputs of the $K$ perceptrons,  
\begin{equation}\label{siebenundzwanzig}  
\sigma = \prod\limits^{K}_{j=1} \; \mbox{sign} \; (\boldsymbol{W}_j \cdot \boldsymbol{S})  
\end{equation}
\begin{figure}
\centerline{
\epsfysize=3cm
\epsffile{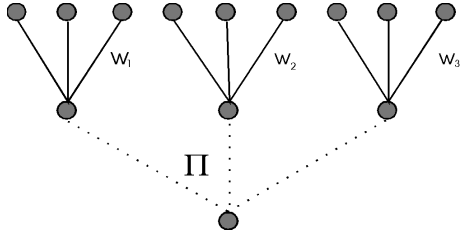}
}
\caption{ Parity machine with a tree architecture. Each of the three weight vectors  
$\boldsymbol{W}_i$ is connected to  only one third of the input vector. The output bit is the product of  
the three hidden units. }
\end{figure}
  
We consider the case, where both of the student and teacher networks are a parity machine   
with the same number of units. The teacher network is presenting a set of examples given by  
\begin{equation}\label{achtundzwanzig}  
\sigma^{\nu}_{B} = \prod\limits^{K}_{j=1} \; \mbox{sign} (\boldsymbol{B}_j \cdot   
\boldsymbol{S}^{\nu}) \; \; (\nu = 1, ..., \alpha N)  
\end{equation}  
The examples should be learned without errors. In this case we are interested in the volume   
$V$ of all student vectors $\{ \boldsymbol{W}_1, ..., \boldsymbol{W}_K \}$ which learn the   
set of examples $\{ (\sigma^{\nu}_{B}, \boldsymbol{S}^{\nu}) \}$ perfectly. $V$ is an   
integral over a $N$--dimensional space and corresponds to the partition sum $Z$ of the   
previous sections. The method of the calculation is similar as before:  
\begin{enumerate}  
\item Average $V$ over all possible sets of inputs $\{\boldsymbol{S}^{\nu}\}$ using the replica   
method.  
\item Calculate the integrals by introducing order parameters $R$ and $Q$, similar to Eqs.   
(\ref{fuenfundzwanzig}) and (\ref{sechsundzwanzig}).  
\end{enumerate}  
In general, an additional average of $V$ over all possible teacher vectors   
$\boldsymbol{B}_i$ is to be performed.  

\begin{figure}
\centerline{
\epsfysize=5cm
\epsffile{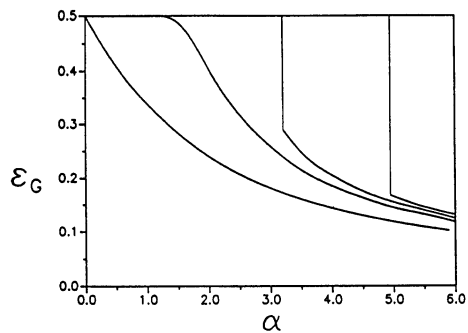}
}
\caption{ Generalization error as a function of the size of the training set for the parity  
machine, which learns perfectly a set of examples given by a teacher parity machine. From left to  
right: $K=1,2,3$ and $4$. (From Opper 1994) }
\end{figure}

The generalization error is shown as a function of the number of
learned examples in Fig.~10 [Opper 1994]. It reveals unexpected
properties of the network: For a large fraction of examples, $ 0 <
\alpha < \alpha_c (K)$, the network cannot generalize at all
$(\varepsilon_g = 1/2)$, although it stores of the order of $N$
examples perfectly! Zero training error does not imply an overlap
between student and teacher network, even for $\alpha >0$.
  
If the number of examples is increased to a critical threshold $\alpha_c (K)$, then the student   
suddenly recognizes the rule, $\varepsilon_g$ jumps to a low value and decreases   
asymptotically as $1/ \alpha$, independently of the number $K$ of hidden units. This property is   
another surprise: The asymptotic behavior is not determined by the Vapnik--Chervonenkis   
dimension, which diverges as $\ln K$ [Barkai et al 1990].  
  
\section{Structure recognition}  
  
Up to now we have discussed supervised learning, that means the input patterns $\boldsymbol{S}^{\nu}$ 
have the labels $\sigma^{\nu}$. But   
there are many applications of neural networks where the labels are not given. In these cases   
the task is to detect a structure in the data [Hertz et al 1991].  
\begin{figure}[h]
\centerline{
\begin{turn}{-90}
\epsfysize=8cm
\epsffile{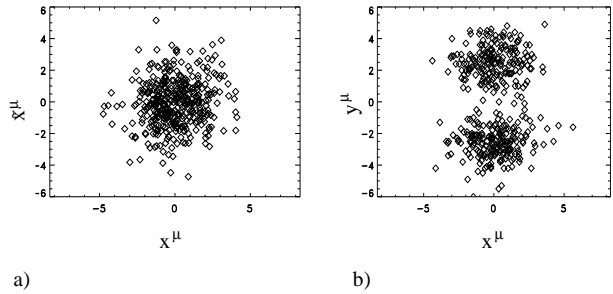}
\end{turn}
}
\caption{ Projection of a distribution of data points. In one direction $\boldsymbol{B}$  
  the overlap has a double peak distribution (b), in all orthogonal
  directions it is Gaussian distributed (a).  (From Biehl, 1997) }
\end{figure}
  
Consider for example two clouds of $\alpha N$ data points as shown in
Fig.~11. The $\boldsymbol{S}^{\nu}$ are distributed in a
$N$--dimensional space according to a mixture of two Gaussian
distributions [Biehl and Mietzner 1993]. This means, that there is a
direction $\boldsymbol{B}$ in data space where the projections
$y_{\nu} = \boldsymbol{B} \cdot \boldsymbol{S}^{\nu} / \sqrt{N}$ have
a double peak distribution. In any  direction $\boldsymbol{W}$
orthogonal to $\boldsymbol{B}$ the corresponding projections $x_{\nu}$
are Gaussian distributed with a single peak, as shown in Fig.~11. Note
that the lengths of all vectors $\boldsymbol{S}^{\nu}, \boldsymbol{W}$
and $\boldsymbol{B}$ are of the order of $N$, while the overlap
$\boldsymbol{S}^{\nu} \cdot \boldsymbol{B}$ is of the order of 
$\sqrt{N}$.
  
Given the $\alpha N$ many data points, we want to find the direction
$\boldsymbol{B}$.  There exists a method, well known in engineering,
which is called ''Principal component analysis" and determines the
direction of maximal variance in data space [Hertz et al 1991]. In
fact there is an algorithm for neural networks which finds this
direction [Oja 1982]. For our example this means, that we want to find a
direction $\boldsymbol{W}$ which minimizes
\begin{equation}\label{neunundzwanzig}  
H(\boldsymbol{W}) = - \sum\limits^{\alpha N}_{\nu = 1} (\boldsymbol{W} \cdot   
\boldsymbol{S}^{\nu} )^2 / N  
\end{equation}  
The minimum of $H$ can be found by calculating the partition sum  
\begin{equation}\label{dreissig}  
Z = \int d^N \boldsymbol{W} \; \delta(\boldsymbol{W}^2 - N) \; \exp (- \beta H   
(\boldsymbol{W}))  
\end{equation}  
in the limit of $\beta \rightarrow \infty$. Here we have again replaced the dynamics of the   
algorithm by a summation over all possibilities. As before we have to average $\ln Z$ over all   
possible data points $\boldsymbol{S}^{\nu}$. The evaluation of the $N$--dimensional integral   
in the limit $N \rightarrow \infty$ yields the order parameter  
\begin{equation}\label{einunddreissig}  
R = \frac{1}{N} \boldsymbol{W} \cdot \boldsymbol{B}  
\end{equation}  
\begin{figure}[h]
\centerline{
\epsfysize=5cm
\epsffile{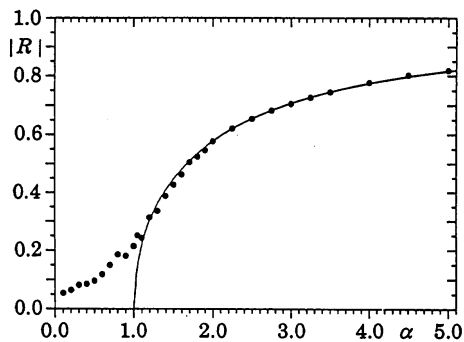}
}
\caption{ Overlap $R$ between the direction $\boldsymbol{B}$ of Fig.~11 and the  
weight vector $\boldsymbol{W}$ of the training algorithm. The theory (solid line) shows a  
transition from zero to nonzero recognition with increasing number of data points in the limit  
$N \rightarrow \infty$. The Monte Carlo simulation (points) for $N =1000$ show that finite size  
effects smoothen the transition. (From Biehl and Mietzner, 1993) }
\end{figure}

The result of this calculation is shown in Fig.~12 [Biehl and Mietzner
1993]. Surprisingly one observes a sharp phase transition. For a small
number of data points the system cannot recognize the direction
$\boldsymbol{B}$ of separation at all. Above a critical number
$\alpha_c$ the symmetry $H(\boldsymbol{W}) = H(-\boldsymbol{W})$ is
spontaneously broken: $|R|$ increases with the deviation from
$\alpha_c$ similar to the magnetization in a ferromagnet.
  
Using the concepts of energy, partition sum and order parameter an unexpected sharp   
transition from zero to good performance was found for the standard method of principal   
component analysis.  
  
\section{Bayesian estimate}  
  
In the previous section we have found the structure of a data distribution by minimizing a cost   
function. However, if one knows something about the structure of the data it is more efficient   
to include this knowledge into the algorithm. Here we want to discuss this problem for the two   
overlapping clouds of data considered in the previous paragraph, see Fig.~11.  
  
The distribution of the data points $\boldsymbol{S}^{\nu}$ is given by  
\begin{eqnarray}\label{zweiunddreissig}  
P(\boldsymbol{S} | \boldsymbol{B} ; \rho) & \propto & \sum\limits_{\tau = \pm 1} \; \exp \left[ 
- \frac{1}{2} (\boldsymbol{S} - \frac{\rho \tau}{\sqrt{N}} \boldsymbol{B} )^2 \right] 
\nonumber\\  
& \propto & \exp [-\beta \; H(\boldsymbol{S}; \boldsymbol{B} , \rho) ]  \; \; (\beta = 1)  
\end{eqnarray}  
This distribution has two parameters: The vector $\boldsymbol{B}$ of length $N$ which gives   
the direction of the cloud separation and the distance $\rho$ between the centers of the clouds.  
  
Now let us assume we know the form of the distribution, Eq.(\ref{zweiunddreissig}), and want to   
estimate its parameters $\boldsymbol{B}$ and $\rho$. Hence our model is for a given distance   
$\tilde{\rho}$:  
\begin{equation}\label{dreiunddreissig}  
P (\boldsymbol{S} | \boldsymbol{W}) \propto \; \exp[- \beta H (\boldsymbol{S};   
\boldsymbol{W}, \tilde{\rho})]  
\end{equation}  
Given the set of data points $\boldsymbol{S}^{\nu}, \nu = 1,..., \alpha N$, the a posteriori   
distribution of directions $\boldsymbol{W}$ is given by the Bayes relation  
\begin{equation}\label{vierunddreissig}  
P(\boldsymbol{W} | \{\boldsymbol{S}^{\nu}\}) = \frac{1}{Z} \; \prod\limits^{\alpha N}_{\nu  
=   
1} \; \delta (\boldsymbol{W}^2 - N) \; \exp [- \beta H (\boldsymbol{S}^{\nu};   
\boldsymbol{W}, \tilde{\rho}  )]  
\end{equation}  
There are several possibilities to estimate a direction $\boldsymbol{W}$ from this distribution   
[Biehl 1997]: Their performance can be measured by the order parameter  
\begin{equation}\label{fuenfunddreissig}  
R = \frac{1}{N} \; \boldsymbol{W} \cdot \boldsymbol{B}  
\end{equation}  
which has a single value in the limit $N \rightarrow \infty$. For example one may maximize the   
a posteriori distribution with respect to $\boldsymbol{W}$. The ''maximum likelihood"   
corresponds to the minimum of  
\begin{equation}\label{sechsunddreissig}  
H (\boldsymbol{W}) = - \sum\limits^{\alpha N}_{\nu = 1} \; \ln \cosh   
\frac{\tilde{\rho}}{\sqrt{N}} \; \boldsymbol{W} \cdot \boldsymbol{S}^{\nu}  
\end{equation}  
which can be studied by calculating $Z$ for $\beta \rightarrow \infty$, using the   
replica method [Barkai and Sompolinsky 1994]. For small cloud separation $\tilde{\rho}$ the maximum   
likelihood solution coincides with first principal component.  
\begin{figure}[h]
\centerline{
\begin{turn}{-90}
\epsfysize=7cm
\epsffile{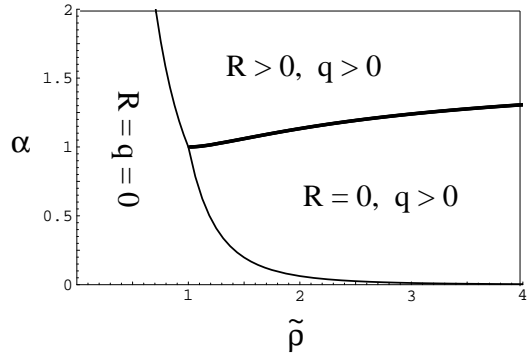}
\end{turn}
}
\caption{ Phase diagram for Bayes estimation of the direction $\boldsymbol{B}$ of the  
two clouds of data points in Fig.~11. $\alpha$ measures the number of data points and $\tilde  
{\rho}$ is the estimated distance of the clouds relative to the true  $\rho = 1$. (From Biehl  
1997) }
\end{figure}
  
Another possibility is to select a direction $\boldsymbol{W}$
according to the a posteriori distribution, Eq.
(\ref{vierunddreissig}), for instance by using the Monte Carlo method
[Watkin and Nadal 1994]. The result is obtained from calculating $Z$
for $\beta = 1$, again by averaging $\ln Z$ over the true distribution
of data points. The result of this Gibbs estimate is shown in Fig.~13 [Biehl 1997].
In the plane of $\alpha$, the size of the data set, and
$\tilde{\rho}$, the estimated distance between the clouds, there are
sharp phase transitions. Recognition $(R > 0)$ appears only for a
sufficiently large number of data points. It is better to use an
estimate $\tilde{\rho}$ which is larger than the true one $\rho = 1$,
since for small $\tilde{\rho}$ the critical fraction $\alpha_c$
diverges as $\tilde{\rho}^{-4}$.
  
\section{On--line training}  
  
In the previous sections {\bf all} of the examples were presented in
the training phase of the neural network. The algorithm used the
training error with respect to all of the examples in order to find
the synaptic weights of the student network. For instance for the
Rosenblatt algorithm, Eq.~(\ref{zwei}) all  examples have to be
predented several times  before the algorithm stops.
  
Now we want to consider a different training algorithm. At each step
only one new example is presented. One does not have to store the
complete set of the examples, but the present weight vector
$\boldsymbol{W}$ is changed due to one new example $(\sigma^{\nu}_B ,
\boldsymbol{S}^{\nu})$. It turns out that such an ''on--line" training
is more efficient in terms of computational effort than the
''off--line" or ''batch" rules of the previous sections, if there are
enough examples available.
  
On--line learning leads to a stochastic differential equation for the
weight vector $\boldsymbol{W}(\nu)$, which becomes a deterministic one
for several order parameters in the limit $N \rightarrow \infty$ [Biehl
and Schwarze 1995, Saad and Solla 1995]. Usually, the dynamics of
on--line learning is not described by a Hamiltonian or a partition
sum, nevertheless there are discontinuous properties as a function of
model parameters.
  
Let us consider a two layer network with continuous output neurons.
The student as well as the teacher network have three hidden units
with weights $\boldsymbol{W}_i$ and $\boldsymbol{B}_i$, the transfer
function of the hidden units is the error function.  For simplicity
the output neuron is linear with fixed weights
\begin{equation}\label{siebenunddreissig}  
\sigma = \sum\limits^3_{i=1} \;\; \mbox{erf} (\boldsymbol{W}_i \; \boldsymbol{S}/\sqrt{2})  
\end{equation}  
The teacher presents $\alpha N$ many examples given by  
\begin{equation}\label{achtunddreissig}  
\sigma^{\nu}_B = \sum\limits^3_{i=1} \;\; \mbox{erf} (\boldsymbol{B}_i \;   
\boldsymbol{S}^{\nu}/\sqrt{2})  
\end{equation}  
The error of a single example is defined as the quadratic deviation  
\begin{equation}\label{neununddreissig}  
\varepsilon (\boldsymbol{W}_1, \boldsymbol{W}_2, \boldsymbol{W}_3,   
\boldsymbol{S}^{\nu}) = \frac{1}{2} (\sigma^{\nu} - \sigma^{\nu}_{B} )^2  
\end{equation}  
In analogy to backpropagation [Hertz et al 1991], the change of
weights is proportional to the gradient of the training error of a
single example:
\begin{equation}\label{vierzig}  
\boldsymbol{W}_k (\nu + 1) = \boldsymbol{W}_k (\nu) - \frac{\eta}{N} \;\;   
\vec{\nabla}_{\boldsymbol{W}_{k}} \;\; \varepsilon  
\end{equation}  
From this equation a system of first order, nonlinear and coupled differential equations can be   
derived for the set of order parameters  
\begin{eqnarray}\label{einundvierzig}  
R_{jk} & = & \frac{1}{N} \; \boldsymbol{W}_j \; \boldsymbol{B}_k \nonumber \\  
Q_{jk} & = & \frac{1}{N} \; \boldsymbol{W}_j \; \boldsymbol{W}_k   
\end{eqnarray}  
  
In our case there are 15 order parameters which change after each presentation of a new   
example. In the limit of $N \rightarrow \infty$ the index $\nu$ becomes a continuous ''time"   
$\alpha$. Hence, one has to calculate the flow of $R_{jk} (\alpha), Q_{jk} (\alpha)$ in the 15   
dimensional space of order parameters which determine the generalization error $\varepsilon_g   
(\alpha)$.   
\begin{figure}[h]
\centerline{
\epsfysize=4cm
\epsffile{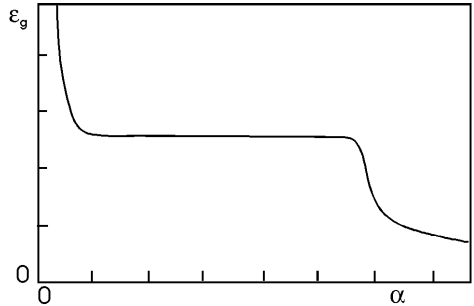}
}
\caption{ Generalization error as a function of the size of training set for a two layer  
network (schematic). The plateau is related to a fixed point with one weak repulsive direction  
for the flow of order parameters. (From Biehl et al 1997) }
\end{figure}
  
It turns out that there are several fixed points of this flow, which have important consequences   
for the behavior of the generalization error. Fig.~14 shows a typical example. For a small   
number of examples, $\varepsilon_g$ decreases fast. But then the generalization error almost   
does not change for a long training period. Suddenly it decreases to good performance of the   
network.  
  
This plateau of $\varepsilon_g (\alpha)$ which is observed in applications, too, can be   
understood in terms of the flow of order parameters [Biehl et al 1997]. There is one fixed point which is   
strongly attractive in almost all directions. But in one or a few directions it has a small repulsive   
component. Hence, the flow remains for quite a while (depending on initial conditions) close to   
this fixed point with a large generalization error, but then it flows away to the completely   
attractive fixed point with zero error.  

The number of fixed points depends on the learning rate $\eta$ of the trainig rule. In our simple example there are at least ten different fixed points for $\eta=1$.   
With increasing $\eta$ some fixed points  split into two, which
usually means that some symmetry is broken. Fixed points  suddenly appear, disappear or
annihilate with varying learning rate $\eta$.  Such discontinuous
behavior is reflected in the generalization error $\varepsilon_g
(\alpha)$.
  
\section{Noise estimation}  
  
The examples, given by a teacher network, may have errors. To what extend can a student   
network derive information about the teacher weights from a set of faulty examples? This   
problem has been investigated in detail [Copelli et al 1997]. We consider a committee machine with a   
tree architecture   
\begin{equation}\label{zweiundvierzig}  
\sigma = \mbox{sign} [ \sum\limits^{K}_{i=1} \;  \mbox{sign} \; \boldsymbol{W}_i \;   
\boldsymbol{S}_i ]  
\end{equation}  
The student as well as the teacher network have the same number $K$ of hidden units. The   
examples are distorted by noise: The bit $\sigma^{\nu}_B$ is flipped with probability 
$\lambda$. For $\lambda = 1/2$ there is   
no information in the examples, but for $0 < \lambda < 1/2$ the student network may obtain an   
overlap to the teacher one with increasing number $\alpha$ of examples.  
  
We study the training algorithm  
\begin{equation}\label{dreiundvierzig}  
\boldsymbol{W}_k (\nu + 1) = \boldsymbol{W}_k (\nu) + \frac{1}{N} F_k
\boldsymbol{S}_k
\end{equation}  
Instead of a parameter $\eta$ we have introduced a function $F_k$
which is determined from a variational principle which maximizes the
decrease of generalization error $\varepsilon_g$ at each training step
[Kinouchi and Caticha 1992]. Hence, one can define an optimal weight
change, which depends on the order parameters. It also contains the noise rate $\lambda$ which is
not known in general; it has to be estimated by a value $\Lambda$,
$F_k (\lambda)$ is replaced by $F_k (\Lambda)$.
\begin{figure}[h]
\centerline{
\begin{turn}{-90}
\epsfxsize=8cm
\epsffile{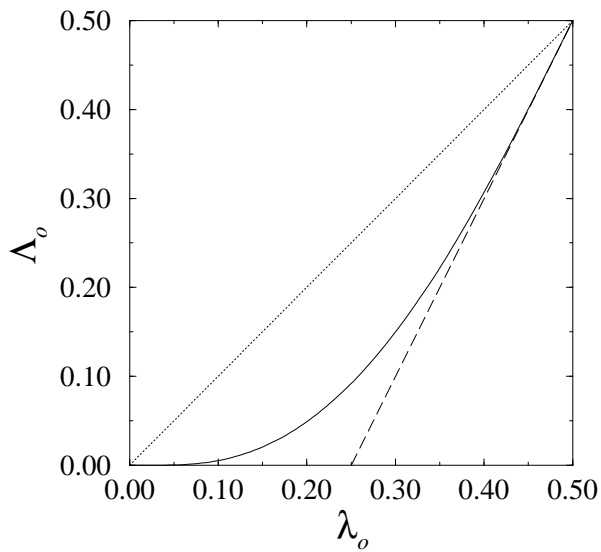}
\end{turn}
}
\caption{ On--line training with optimal weights. $\Lambda$ is the estimated and  
$\lambda$ the true noise level of the training examples. Above the solid line perfect  
generalization is possible, $\varepsilon_g (\alpha \rightarrow \infty) = 0$. Below the dashed  
line the network cannot generalize at all. Between the dashed and solid line a partial  
recognition of the teacher network is possible. (From Copelli et al, 1997) }
\end{figure}
  
The generalization error has been calculated for optimal on--line learning [Copelli et al 1997],   
Fig.~15 shows the result. There are sharp boundaries in the $(\lambda, \Lambda)$ plane where   
the behavior of the network changes drastically. Fastest decay of $\varepsilon_g (\alpha)$ is   
obtained if the true noise is estimated correctly, $\lambda = \Lambda$, as expected. If the   
estimated noise parameter $\lambda$ is large enough, then $\varepsilon_g (\alpha)$ still   
decreases to zero in the limit of an infinite number $\alpha$ of examples. However, if 
$\Lambda$   
is below the dashed line, then the network cannot generalize at all. If $\Lambda$ lies in 
the   
intermediate region then the generalization error decreases to a nonzero value; the network can   
generalize only partially. Again we observe sudden changes in the behavior of the network as 
a function of model parameters.  
  
\section{Time series generalization}  
  
Most of the work on the statistical physics of neural networks has been done on static data. A   
set of input vectors $\{\boldsymbol{S}^{\nu} \}$ is taken from a static distribution and   
classification labels $\{ \sigma^{\nu}\}$ are taken from a static rule. Only recently this research   
program has been extended to the analysis of time series [Eisenstein et al, 1995], which is an 
important field of   
applications of neural networks. 

\begin{figure}[h]
\centerline{
\epsfysize=2cm
\epsffile{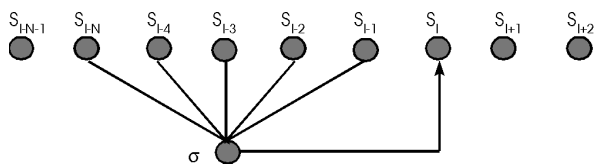}
}
\caption{ A perceptron working as sequence generator. The sequence $S_l$ of numbers  
is generated by a perceptron moving to the right. }
\end{figure}
 
In the simplest case our elementary unit, the perceptron, is trained to a sequence $S_i$ of real  
numbers, where $i$ is a discrete time. As shown in Fig.~16, the perceptron takes a window of  
$N$ numbers as input and makes a prediction of the following number, 
\begin{equation}\label{vierundvierzig} 
S_l^{\prime} = \tanh [ \frac{\beta}{N} \; \sum\limits^N_{j = 1} \; W_j \; S_{l - j} ] 
\end{equation} 
In the training phase the weights $\boldsymbol{W}$ are changed to decrease the error 
$(S_l^{\prime} - S_l)^2 $. 
 
In order to apply the concepts of statistical physics one needs a well defined sequence  
$\{S_i\}$. As before it may be given by a teacher perceptron with weight vector  
$\boldsymbol{B}$. For a given window of $N$ numbers $(S_{l-N, ... , } S_{l-1})$ the  
perceptron defines the number $S_l$. Then it moves one step and generates $S_{l+1}$. It  
turns out that the generation of time series is already an interesting problem with many  
unsolved puzzles [Eisenstein et al 1995, Kanter et al 1995, Schr\"oder and Kinzel 1997]. 
 
The numerical investigation of the sequence generator shows that an initial state of  
Eq~(\ref{vierundvierzig}) approaches a quasi periodic attractor which is related to a peak in the 
Fourier  
spectrum of $\boldsymbol{W}$ [Eisenstein et al 1995]. Hence, the perceptron selects one 
mode of the  
couplings. Therefore it is useful to study couplings with a single Fourier component 
\begin{equation}\label{fuenfundvierzig} 
W_j = \cos (2 \pi k \; \frac{j}{N} - \pi \phi ) 
\end{equation} 
$k$ is the frequency and $\phi$ the phase of the weights. An attractor of  
Eq.~(\ref{vierundvierzig}) is the solution of 
\begin{equation}\label{sechsundvierzig} 
S_l = \tanh [ \frac{\beta}{N} \; \sum\limits^N_{j=1} \, \cos (2\pi k \frac{j}{N} - \pi \phi) S_{l- 
j}] 
\end{equation} 
Recently this equation could be solved analytically [Kanter et al 1995]: For small values of 
$\beta$ the  
attractor is zero, $S_l = 0$. For a critical value, which is independent of the frequency $k$,  
\begin{equation}\label{siebenundvierzig} 
\beta_c = 2 \frac{\pi \phi}{\sin \pi \phi} 
\end{equation} 
there appears the solution 
\begin{equation}\label{achtundvierzig} 
S_l = \tanh [A(\beta) \; \cos(2\pi (k + \phi) \frac{l}{N} )] 
\end{equation} 
The phase $\phi$ of the weights shifts the frequency $k + \phi$ of the solution. The amplitude  
$A(\beta)$ is a continuous function of the distance $\beta - \beta_c > 0$ to the critical point.  
For $\beta \rightarrow \infty \;\; \; \tanh$ is replaced by sign and the sequence generator  
becomes a bit generator. In this case the solution, Eq.~(\ref{achtundvierzig}), is more  
complicated, but again the phase $\phi$ shifts the frequency of the bit sequence [Schr\"oder 
and Kinzel 1997]. 
\begin{figure}[h]
\centerline{
\epsfxsize=8cm
\epsffile{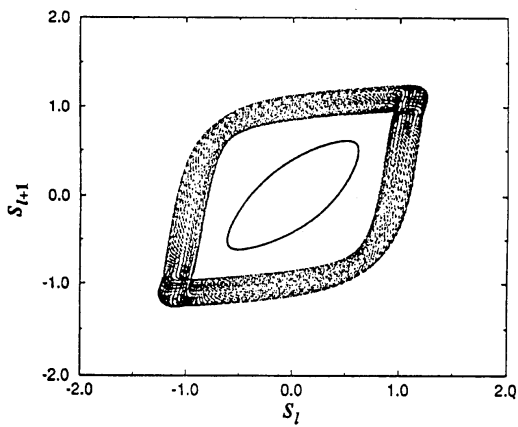}
}
\caption{ Return map of a sequence generated by a multilayer network with two hidden  
units. The one--dimensional attractor in the middle becomes a two--dimensional one, if the  
parameter $\beta$ is increased. (From Kanter et al, 1995) }
\end{figure}
 
If the network is a multilayer perceptron with $K$ hidden units, then
the attractor of the sequence generator is a kind of superposition of
the single modes of each hidden unit [Kanter et al 1995].  Each unit
has ist own critical point and the number of nonzero solution
determines the dimension of the attractor. This is shown in the return
map of Fig.~17 where $S_{l+1}$ is plotted against $S_l$
for $K=2$ hidden units. Increasing $\beta$ first one component is
activated leading to a one dimensional attractor shown in the middle
of the figure. Since in general the frequency is irrational, the
attractor is a continuous curve. For larger values of $\beta$ the
second unit is activated, giving the two--dimensional attractor with
larger amplitude.
 
\section{Summary} 
 
Concepts of statistical physics have successfully been applied to the theory of neural  
computation. The cooperative behavior of a large number of interacting neurons can be  
described in terms of partition sums and order parameters. The competition between training  
error and entropy may lead to discontinuous properties.\\ 
The approach of statistical mechanics has several advantages: 
\begin{enumerate} 
\item  Networks with an infinite number of neurons and synapses can be calculated analytically.  
Complex cooperative behavior of interacting neurons is described in terms of a few order  
parameters. 
\item The results are obtained for a typical situation, for instance for the most general set of  
examples. 
\item One obtains exact mathematical relations between the observed cooperative properties of  
the network, its model parameters and the size of the training set. 
\item Many networks and algorithms show discontinuous properties as a function of model  
parameters or the number of presented examples. Statistical physics can describe such sudden  
changes in the cooperative behavior of the network. 
\end{enumerate} 
Statistical mechanics of neural networks has been applied to a variety of problems; we just  
want to mention learning from examples, generalization, associative memory, attractor  
networks, structure recognition, clustering, classification, coding and time series analysis. For  
all of these problems general properties have been calculated mathematically. Novel and  
unexpected results have been found. Hence, I think that in the last 15 years theoretical physics  
has successfully contributed to our understanding of neural networks, with impact on  
neurobiology and computer science. 
    
{\bf Acknowledgement:} The author thanks Michael Biehl for comments on the manuscript.  
       
{\large \bf Literature}     
       
Hertz, J.~A., A.~Krogh and R.~G.~Palmer, 1991, Introduction to the Theory of Neural      
Computation (Addison Wesley)    
     
Amit, D.~, H.~Gutfreund and H.~Sompolinsky, 1987, Annals of Physics, {\bf 173}, 30-67    
     
Barkai, E.~, D.~Hansel and I.~Kanter, 1990, Phys.~Rev.~Lett.~{\bf 65} 2312    
     
Barkai, N.~and H.~Sompolinsky, 1994, Phys.~Rev.~{\bf E 50}, 1766    
     
Biehl, M.~and A.~Mietzner 1993, Europhys.~Lett.~{\bf 24}, 421--426    
     
Biehl, M.~and H.~Schwarze, 1995, J.~Phys.~ {\bf A 28}, 643(?)    
     
Biehl, M.~, 1997, lecture notes, University of W\"urzburg   
     
Biehl, M.~, P.~Riegler and C.~W\"ohler, 1997, to be published    
     
Copelli, M.~, R.~Eichhorn, O.~Kinouchi, M.~Biehl, R.~Simonetti, P.~Riegler and N.~Caticha,     
1997, Europhys.~Lett.~{\bf 37}, 427-432    
     
Eisenstein, E.~, I.~Kanter, D.~Kessler and W.~Kinzel, 1995, Phys.~Rev.~Lett.~{\bf  74}, 6    
     
Fischer, K.~H.~and J.~A.~Hertz, 1991, Spin Glasses, (Cambridge University Press)    
     
Gardner, E.~, 1988, J.~Phys.~{\bf A 21}, 257--270

Hopfield, J.~J.~, 1982, Proceedings of the National Academy of Sciences, USA {\bf 79},      
2554--2558    
   
Kanter, I.~, D.~A.~Kessler, A.~Priel and E.~Eisenstein,  1995, Phys.~Rev.~Lett.~{\bf 75},     
2614--2617    

Kanter, I. and H. Sompolinsky, 1987, Phys. Rev. {\bf A 35}, 380

Kinzel, W. , 1985, Z. Physik {\bf B60}, 205
     
Kinouchi, O.~and N.~Caticha, 1992, J.~Phys.~{\bf A 25}, 6243    
     
Oja, E.~, 1982, J.~Math.~Biol.~{\bf 15}, 267    
     
Opper, M.~, 1994, Phys.~Rev.~Lett.~{\bf 72}, 2113    
     
Opper, M.~ and W.~Kinzel, 1996, in Models of Neural Networks III, ed.~by E.~Domany,      
J.~L.~van Hemmen and K.~Schulten (Springer, Berlin)    
     
Saad, D.~ and S.~A.~Solla, 1995, Phys.~Rev.~Lett. {\bf 74}, 4337    
     
Schl\"afli, L.~, 1852, Theorie der vielfachen Kontinuit\"at, Gesammelte Mathematische      
Abhandlungen, ed.~Steiner--Schl\"afli--Komittee Basel, Birkh\"auser p 171    
     
Schr\"oder, M.~and W.~Kinzel, 1997, to be published    
     
Schwarze, H.~, M.~Opper and W.~Kinzel, 1992, Phys.~Rev.~{\bf A 46}, R 6185    
     
Seung, H.~S.~, H.~Sompolinsky and N.~Tishby, 1992, Phys. Rev. {\bf A 45}, 6056    
     
Watkin, T.~L.~H.~, A.~Rau and M. Biehl, 1993, Rev.~ Mod.~Phys.~ {\bf 65}, 499    
     
Watkin, T.~L.~H.~and J.~--P.~Nadal, 1994, J.~Phys.~{\bf A 27}, 1889    
     
Yeomans, J.~, 1992, Statistical mechanics of phase transitions, (Clarenden Press, Oxford)

\end{document}